\documentclass[11pt]{article}

\usepackage[utf8]{inputenc}
\usepackage[T1]{fontenc}
\usepackage{amsmath,amssymb}
\usepackage{graphicx}
\usepackage{booktabs}
\usepackage[hypertexnames=false]{hyperref}
\usepackage{xurl}  
\usepackage{xspace}
\usepackage{microtype}
\usepackage{xcolor}
\usepackage[margin=1in]{geometry}

\hypersetup{
    colorlinks=true,
    linkcolor=blue,
    urlcolor=cyan,
    citecolor=blue,
    pdftitle={Needles at Scale: LLM-Assisted Target Selection for Windows Vulnerability Research},
    pdfauthor={Michael J.\ Bommarito II},
    pdfsubject={Vulnerability research, binary analysis, LLM, target selection, Windows},
    pdfkeywords={vulnerability research, target selection, binary analysis, LLM enrichment, Windows, attack surface, prioritization}
}

\newcommand{\nfns}{7{,}231{,}419\xspace}
\newcommand{\nfnsM}{7.23M\xspace}
\newcommand{\nbins}{5{,}888\xspace}
\newcommand{\nnamed}{5{,}552{,}861\xspace}
\newcommand{\nnamedM}{5.55M\xspace}
\newcommand{\sys}{\textsc{Symbolicate-Enrich-Sample}\xspace}
\newcommand{\glaurung}{\texttt{glaurung}\xspace}
\newcommand{\code}[1]{\texttt{#1}}

\title{\textbf{Needles at Scale}:\\
  LLM-Assisted Target Selection\\
  for Windows Vulnerability Research}

\author{
Michael J.\ Bommarito II\thanks{Portions of this work were prepared with
assistance from large language models. The author is solely responsible for
all content, including any errors or omissions. This work was conducted for
defensive and authorized vulnerability-research purposes; see the data-release
and ethics notes in Section~\ref{sec:discussion}.}\\
\texttt{michael.bommarito@gmail.com}
}

\date{May 2026}

\begin{document}
\maketitle

\begin{abstract}
The attack surface of a modern operating system is a haystack: thousands of
signed binaries and millions of functions, almost none relevant to any given
vulnerability. A human analyst or an LLM agent must pick the function worth
reading before analyzing it. At whole-OS scope, this \emph{target selection},
not the analysis, is the binding constraint. We present \sys, a low-cost batch
pipeline that turns a corpus of production Windows binaries into a queryable,
priority-ranked research queue. We (i)~recover function-level symbols for
stripped vendor binaries by auto-fetching the public symbol files and joining
them to a recovered call graph; (ii)~attach cheap, deterministic structural
features to each named function and, conditioned on those features, use a
low-cost language model to assign a reachability tier, a risk level, a
bug-class hypothesis, and a rationale; and (iii)~draw diverse, prioritized
batches via a priority-weighted importance sampler. The contribution is a
\emph{selection substrate}: the prioritization layer a downstream detector or
LLM agent runs on top of. Across a whole Windows image of \nfns\ functions, the
labels are markedly selective, and stacking deterministic filters on them leaves
a ${\sim}22$K-function shortlist: the candidate needles, few enough for a human
or agent to work through. We characterize the
pipeline's selectivity and its failure modes, describe the methodology, and
report aggregate statistics; we withhold the derived dataset for legal and
dual-use reasons.
\end{abstract}

\section{Introduction}
\label{sec:intro}

Finding a memory-safety bug in a closed-source operating system is two problems
wearing one coat. The visible problem is \emph{analysis}: reasoning about a
function's bounds, lifetimes, and reachability until one can exhibit a primitive
or cite the construct that makes it safe. The prior, hidden problem is
\emph{target selection}: out of the millions of functions that ship in a
Windows image, deciding which few hundred deserve that expensive analysis at
all. Analysis has attracted enormous tooling: symbolic execution, decompilers,
fuzzers. Selection is still largely done by intuition, reputation
(``SMB is juicy''), and \code{grep}.

This imbalance sharpens as analysis is delegated to large language model (LLM)
agents, which can already find real bugs when aimed at a specific
target~\cite{bigsleep}. An agent that reads one decompiled function competently
still must be \emph{pointed} at the right function; aimed at a random one it
burns a costly context window confirming that C-runtime startup code is not an
attack surface.
The bottleneck moves from ``can we analyze this function'' to ``which function,
of millions, next.'' Recent work makes the same diagnosis: that vulnerability
research is resource-constrained prioritization, and reframes
selection as information retrieval or as candidate ranking with
LLMs~\cite{siftrank,onthemoney,promptingpriorities}. We take that framing as our
starting point and ask a complementary, systems question: \emph{what does a
selection substrate look like when it must cover an entire production operating
system (millions of real, stripped, vendor-shipped functions) cheaply enough
to run end to end?}

Our answer is \sys, a three-stage batch pipeline whose only job is to order
the expensive reads (Section~\ref{sec:method}). The contribution is not a bug
detector; it is the prioritization layer a detector should run on top of, built
and characterized at production Windows scale. Windows is not incidental: the
\emph{Symbolicate} stage depends on Microsoft's public symbol server, which
publishes function names for stripped \emph{system} binaries, an unusually
generous arrangement. The \emph{Enrich} and \emph{Sample} stages are
platform-independent, so the pipeline transfers to any ecosystem with a
comparable public symbol source. Linux distributions provide one through
\code{debuginfod}; platforms that do not publish system symbols (notably macOS)
would need a different naming step.

\paragraph{Contributions.}
\begin{itemize}
  \setlength{\itemsep}{1pt}
  \item A reproducible method to recover function-level symbols for stripped
  production Windows binaries at scale, by auto-fetching vendor PDBs and
  joining them to a recovered call graph (\S\ref{sec:symbolicate}).
  \item \emph{Feature-grounded} LLM enrichment: each function is labelled with
  reachability, risk, bug-class hypothesis, and rationale conditioned on
  cheap deterministic features, at low per-function cost
  (\S\ref{sec:enrich}).
  \item A priority-weighted importance sampler that converts the enriched
  corpus into a diverse, prioritized queue (\S\ref{sec:sample}).
  \item A selectivity and failure-mode characterization of the resulting
  pipeline on a whole-OS Windows corpus (\S\ref{sec:eval}).
  \item A discussion of the legal and dual-use reasons we publish the method
  and aggregate statistics but withhold the derived dataset
  (\S\ref{sec:discussion}).
\end{itemize}

\paragraph{Related work, in brief.}
LLM-based vulnerability detection has largely targeted \emph{source} or
synthetic/open-source-compiled code at the scale of thousands to tens of
thousands of labelled samples, optimizing detection
accuracy~\cite{llmvulnsurvey,secvuleval}. The prioritization
turn (selection as the real bottleneck) appears in SiftRank's
information-retrieval framing~\cite{siftrank}, in scaling-oriented candidate
ranking~\cite{onthemoney}, and in triage-evaluation
studies~\cite{promptingpriorities}, and scalable Windows-specific hunting has
been demonstrated for individual subsystems such as
ALPC~\cite{needlealpc}. The stance is older than LLMs: actionable-alert
identification has ranked static-analysis warnings to cut false-positive triage
cost since well before LLMs~\cite{heckman2011} and remains
active~\cite{awisurvey}, and directed greybox fuzzing steers scarce
dynamic exploration toward reachable targets~\cite{aflgo}; we bring the same
prioritization stance to symbol-level selection over production binaries at
corpus scale. On the corpus side, public Windows binary
\emph{indices} provide metadata and download links but no function-level
symbolication or enrichment~\cite{winbindex}. To our knowledge no public
artifact combines production Windows scale, PDB-symbolicated function
granularity, and feature-grounded LLM risk/reachability enrichment; that
combination, and its selectivity, is what we report.

\section{Method}
\label{sec:method}

\begin{figure}[t]
  \centering
  \includegraphics[width=\textwidth]{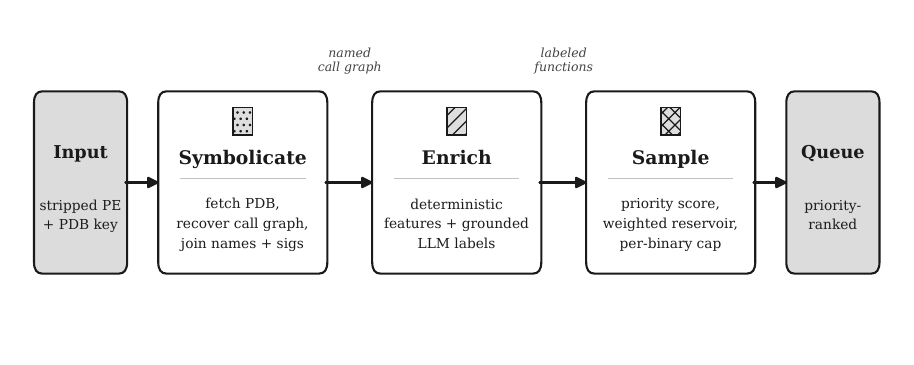}
  \caption{The \sys\ pipeline. \emph{Symbolicate} recovers a named call graph
  from a stripped PE via its PDB; \emph{Enrich} attaches deterministic
  structural features and feature-grounded LLM labels; \emph{Sample} draws a
  priority-ranked queue. Only aggregate structure crosses each stage; the
  expensive model sees a compact feature summary, never decompiled bytes.}
  \label{fig:pipeline}
\end{figure}

\sys\ runs as three batch stages over a binary corpus
(Figure~\ref{fig:pipeline}), writing to a single relational store keyed by
\code{(binary\_sha256, function\_va)}. We summarize each; the pipeline is fully
reproducible from binaries that are themselves publicly retrievable from
Microsoft-hosted symbol and file endpoints under Microsoft's terms.

\subsection{Symbolicate}
\label{sec:symbolicate}
Shipping Windows PE files carry no local symbols, but each contains a CodeView
(RSDS) record naming a PDB and a \code{<GUID><age>} key. We parse that record,
construct the canonical symbol-server URL, and fetch and cache the PDB. A
decompiler pass (\glaurung\footnote{\glaurung\ is our own binary-analysis
tooling; the method is agnostic to the specific decompiler.}) recovers the
function set and inter-procedural call graph; we then join Microsoft's public
function names onto recovered function addresses and join a catalog of
${\sim}20$K Win32/WDK API prototypes onto called imports to recover the
\emph{callee} prototypes for each call site (we do not recover full target-%
function type signatures). Functions with no public name remain as
synthetic \code{sub\_<addr>} entries and are excluded from enrichment. This
stage is deterministic, cacheable, and I/O- rather than model-bound.
The function names themselves are published by the vendor;
recovering them is symbolication, not decompilation of secret logic.

\subsection{Enrich}
\label{sec:enrich}
Enrichment attaches a \emph{grounding feature vector} and then a \emph{model
label} to each named function.

\paragraph{Deterministic features.} From the call graph and disassembly
we compute, per function: the exported flag; caller/callee counts and the
PDB-resolved names of the top callers and callees; whether the function calls a
risk-relevant memory or user-buffer API, separated into copy/write sinks
(\code{memcpy}, \code{memmove}, \code{RtlCopy*}, \code{memset}), pool
allocators, and user-pointer probes; basic-block count and
size; and a breadth-first \emph{reach depth} from entry points (exports,
dispatch routines, graph roots). These features are cheap, exact, and carry
much of the real signal: a function with no observed copy/write sink is less
likely to host the copy-sink bug classes this filter targets, though it may
still contain other defects.

\paragraph{Model label.} We prompt a low-cost LLM, served at a
discounted batch tier, with the function's name, recovered callee prototypes,
and \emph{the grounding features}, and require a structured output: a reachability
tier (\code{remote}, \code{local-ipc}, \code{local-user}, \code{admin-only},
\code{kernel-internal}, \code{library-internal}); a risk level
(\code{critical}/\code{high}/\code{medium}/\code{low}/\code{info}); a bug-class
hypothesis; a confidence; a one-line role; and a free-text rationale. These
tiers are \emph{model-assigned prioritization labels, not proofs}: reach depth
and names are evidence, but indirect calls, RPC routing, configuration, and
runtime guards can move the true boundary. The model is instructed to reason
\emph{from the supplied features} (citing the copy primitive, the caller
names, the reach depth) rather than from the name alone.
Functions matching known runtime/CRT-helper patterns are labelled by rule
(forced to \code{library-internal}/\code{info}) and never sent to the model,
removing a large, predictable, low-value population from the bill. The job is
sharded by a hash of binary identity across workers, each batching functions per
model call with bounded concurrency, so millions of functions complete in hours.
The expensive component only ever sees a compact structured summary, never raw
decompiled bytes.

\subsection{Sample}
\label{sec:sample}
The enriched store is queried by a priority-weighted importance sampler. We
score each function with a hand-tuned \emph{priority score} (a heuristic
ordering, not a calibrated probability)
\begin{equation}
\mathrm{priority} \;=\; w_{\text{reach}}\cdot w_{\text{risk}}\cdot
w_{\text{conf}}\cdot \text{bonus},
\end{equation}
where the weights map the model's tiers to monotone numeric values and
\code{bonus} multiplicatively rewards \emph{deterministic} evidence the model
cannot fabricate (actually calling a copy/write sink, being exported,
sitting near an entry point, a parser-sized body), while penalizing trivial
stubs. Given a filter (e.g.\ ``remote, critical-or-high, calls a copy
primitive''), the sampler draws without replacement via a weighted reservoir
(Efraimidis--Spirakis, key $u^{1/\mathrm{priority}}$~\cite{efraimidis}), so a session
covers the high-value space with \emph{diversity} rather than re-returning the
same top-$k$; a per-binary cap prevents one module from dominating. A
\code{top-N} mode serves the sharpest leads directly. The sampler reads the
store read-only and is safe to run while enrichment is still in progress.

Which regime fits depends on the campaign. Broad, fuzzing-style sweeps favor the
weighted draw, which spreads effort across the high-value space; targeted,
hypothesis-driven work favors a tight filter with \code{top-N}, concentrating on
one bug class, binary, or reachability tier. These sit on a spectrum from
coverage to precision, and the priority score supports other points on it:
stratified quotas per binary or bug class, threshold cuts, or re-weighting the
features toward an analyst's current interest. The scoring fixes an ordering;
the choice of how to draw from it is left to the caller.

\section{Evaluation}
\label{sec:eval}

We are explicit about what we can and cannot evaluate. We \emph{cannot} report a
true-positive rate: establishing ground truth requires the very decompile-level
verification the pipeline exists to prioritize, and at \nnamedM\ functions no
exhaustive oracle exists. We \emph{can} evaluate the properties that determine
whether the layer usefully concentrates attention (\textbf{selectivity} and
distributional \textbf{conservatism}) and what it costs.

\paragraph{Corpus.} We apply \sys\ to \nbins\ x86-64 Windows~11 binaries at
build 10.0.26100, the servicing base shared by the 24H2 and 25H2 feature
updates, drawn from two installations at 2025 patch levels
(Table~\ref{tab:corpus}): \nfns\ recovered functions, of which $76.8\%$
received public PDB names and were enriched; the rest are unnamed
\code{sub\_*} stubs.

\begin{table}[t]
  \centering
  \small
  \begin{tabular}{rrr}
    \toprule
    \textbf{Corpus quantity} & \textbf{Count} & \textbf{Share} \\
    \midrule
    Signed PE binaries            & 5{,}888       &           \\
    Functions recovered           & 7{,}231{,}419 &           \\
    PDB-named functions           & 5{,}553{,}074  & 76.8\%    \\
    of which enriched             & 5{,}552{,}861  & 99.996\%  \\
    Unnamed \code{sub\_*} stubs   & 1{,}678{,}345  & 23.2\%    \\
    \bottomrule
  \end{tabular}
  \caption{The corpus: a whole production Windows image. \textbf{Share} is of all
  recovered functions, except the indented \emph{enriched} row, which is of the
  PDB-named functions. Only PDB-named functions are enriched; unnamed stubs carry
  too little signal to label and are excluded. Enrichment reached all but 213
  named functions.}
  \label{tab:corpus}
\end{table}

\subsection{Selectivity and conservatism}
\begin{figure}[t]
  \centering
  \includegraphics[width=\textwidth]{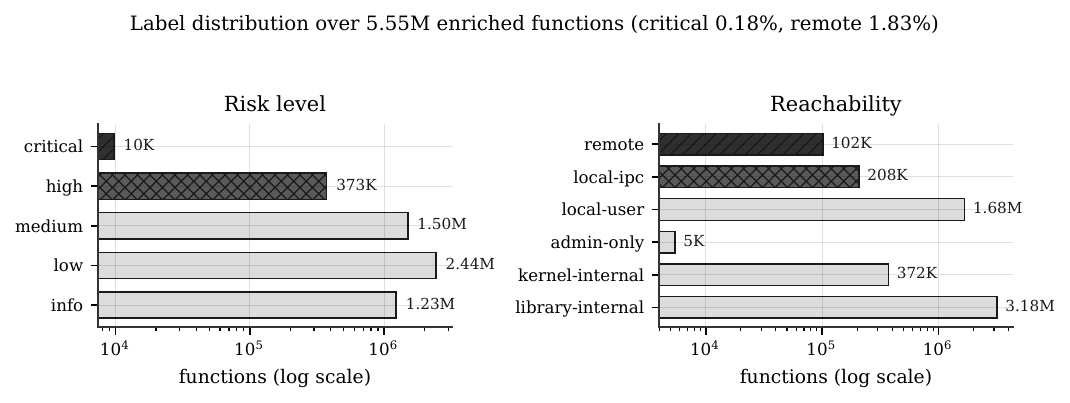}
  \caption{Label distribution over \nnamed\ enriched functions (log $x$).
  The classifier is conservative: \code{critical} and \code{remote} (the tiers
  the sampler targets) are each a small fraction of the corpus, while the bulk
  of functions are labelled \code{library-internal}; manual spot checks of that
  bulk found the expected runtime/boilerplate code.}
  \label{fig:dist}
\end{figure}

Over the \nnamed\ enriched functions the label distribution forms a steep,
conservative pyramid rather than collapsing toward the alarming end
(Figure~\ref{fig:dist}). The model reserves \code{critical} for $0.18\%$ of
functions and the \code{remote}-facing tier for $1.83\%$; a majority land in
\code{library-internal}, and manual spot checks found CRT startup, C++ exception
machinery, and other boilerplate pushed to \code{info}/\code{low}.
This is precisely the population an unaided researcher wastes time
re-dismissing. In sampled cases the rationales appear \emph{grounded}: the
explanations cite the supplied features (``\code{reach\_depth} 48'', ``calls
\code{memcpy}'', ``depth 1 from the dispatch entry'') rather than generic
boilerplate, which is what makes the subsequent human read fast.
Table~\ref{tab:examples} makes this concrete on four recognizable Win32/NT APIs:
\code{wcscpy} is flagged because it calls \code{memcpy}, while \code{memcpy}
itself is demoted to \code{info}, and the exported \code{VirtualProtect} is
demoted as a thin syscall wrapper rather than flagged on its name.

\begin{table}[t]
  \centering
  \footnotesize
  \setlength{\tabcolsep}{5pt}
  \begin{tabular}{@{}lcccccccc@{}}
    \toprule
    & \multicolumn{5}{c}{\textit{Deterministic features}}
    & \multicolumn{3}{c}{\textit{Model label}} \\
    \cmidrule(lr){2-6}\cmidrule(lr){7-9}
    Function \scriptsize(binary)
      & exp. & call. & blk. & dep. & sink
      & reachability & risk & bug-class hyp. \\
    \midrule
    \code{RtlDecompressBuffer} \scriptsize(ntdll)
      & yes & 0 & 6 & 0 & none
      & local-user & high & int-overflow \\
    \code{wcscpy} \scriptsize(ntdll)
      & yes & 1 & 1 & 1 & \code{memcpy}
      & local-user & high & oob-write \\
    \code{VirtualProtect} \scriptsize(KernelBase)
      & yes & 0 & 7 & 0 & none
      & local-user & low & none-obvious \\
    \code{memcpy} \scriptsize(ntdll)
      & yes & 575 & 37 & 1 & none
      & library-internal & info & none-obvious \\
    \bottomrule
  \end{tabular}
  \caption{Four enriched records for recognizable Win32/NT APIs.
  \emph{exp.}\,=\,exported, \emph{call.}\,=\,in-binary callers,
  \emph{blk.}\,=\,basic blocks, \emph{dep.}\,=\,reach depth,
  \emph{sink}\,=\,copy/write sink. The
  labels track the evidence, not the name: \code{wcscpy} is flagged because it
  calls \code{memcpy}, while \code{memcpy} itself (the primitive, with 575
  callers) is demoted to \code{info}; the exported \code{VirtualProtect} is
  demoted as a thin syscall wrapper with no copy/parse sink, whereas
  \code{RtlDecompressBuffer} is flagged on role (size arithmetic in
  decompression). Labels are model outputs, not ground truth.}
  \label{tab:examples}
\end{table}

\begin{figure}[t]
  \centering
  \includegraphics[width=\textwidth]{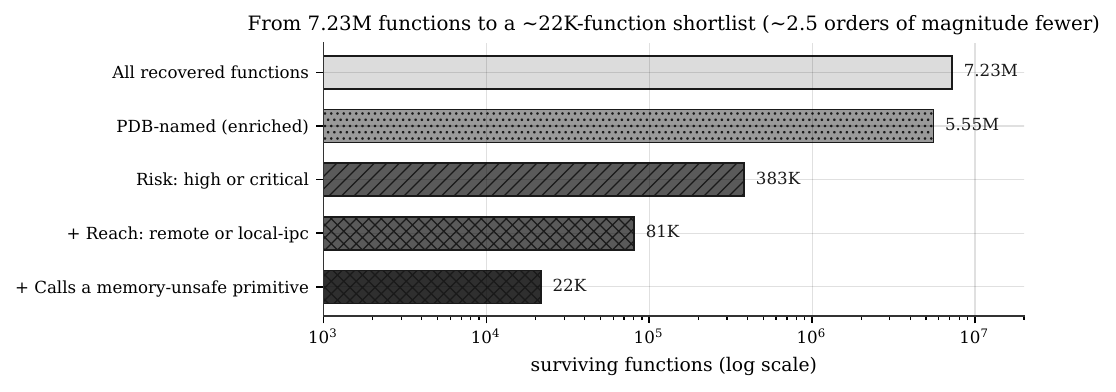}
  \caption{Selectivity funnel. Each successive filter is a column in the
  enriched store; the combination reduces the search space by ${\sim}2.5$
  orders of magnitude to a ${\sim}22$K-function shortlist of candidates.}
  \label{fig:funnel}
\end{figure}

\paragraph{Selectivity funnel.} Stacking the model's labels with the
deterministic features compounds the reduction (Figure~\ref{fig:funnel}):
restricting to high/critical risk \emph{and} remote-or-local-IPC reachability
\emph{and} a call to a risk-relevant memory or user-buffer API narrows
\nfnsM\ functions
to ${\sim}22$K (${\sim}2.5$ orders of magnitude), and the \code{critical}
$\cap$ \code{remote} corner to ${\sim}2{,}100$. We stress these are
\emph{candidate} counts, not confirmed vulnerabilities; the value of the
shortlist is that it is small enough for a human or agent to work through.

\subsection{Failure modes}
Sampling the top tier exposes two systematic weaknesses. (1)~\textbf{Top-tier
over-reach}: within the small \code{critical} bucket the model occasionally
assigns a \emph{specific} bug class (use-after-free, unchecked-pointer-deref) to
functions whose feature vector contains \emph{no} copy or memory-write
sink---it names a wound with no weapon present. (2)~\textbf{Reachability
inflation on deep parsers}: it will label a structurally parser-like function
\code{remote} even at a call-graph reach depth of $40$+, assuming
network-sourced input the features do not establish. We saw both errors in the
sampled top tier; the first is partly \emph{mechanically} detectable: for
copy-oriented bug classes, a concrete bug-class label with no corresponding sink
in the features is unsupported (for other classes it is merely weakly supported,
not impossible). A simple guard (forbid a concrete copy-class label without a
sink, and down-weight \code{critical} rows lacking a risky primitive in the
sampler) mitigates the first failure mode without re-running the model.

\subsection{Cost}
Low cost is a design goal, not an afterthought. Because the model component
consumes only a compact structured feature summary (not decompiled code), and a
large predictable fraction of functions is excluded by rule before any model
call, per-function spend is small: the full \nfns-function corpus is enrichable
with a low-cost model at batch pricing, inexpensively relative to manual triage
of the same surface. The deterministic features, which carry much of the
ranking signal, cost nothing beyond the decompiler pass.

\section{Discussion}
\label{sec:discussion}

\paragraph{Why we release the method but not the data.} The enriched corpus is
derived from Microsoft's copyrighted binaries. Redistributing function-level
derived analysis at scale may create copyright or license risk that a
methodology description and aggregate counts do not. (The function
\emph{names} are independently published by Microsoft's symbol server, but the
derived risk/reachability analysis is our work \emph{about} protected binaries.)
The closest public artifact, a Windows binary \emph{index}, is deliberately
metadata-only and links back to Microsoft's own server~\cite{winbindex}; we
adopt the same posture and therefore avoid redistributing derived
function-level annotations. Anyone with the binaries (publicly retrievable
from Microsoft-hosted symbol and file endpoints under Microsoft's
terms) can reproduce the corpus from Section~\ref{sec:method}.

\paragraph{Dual use.} A ranked, remotely-reachable attack-surface map is an
offense--defense-symmetric artifact: published openly with no patch in hand, it
uplifts attackers at least as much as defenders, a concern that sharpens as
LLM agents become able to act on such targets autonomously~\cite{fang2024oneday}. This is an independent
reason, beyond copyright, to withhold the dataset and to frame the
contribution as a \emph{prioritization methodology} whose intended use is to
accelerate responsible vulnerability research and coordinated disclosure.

\paragraph{Limitations and threats to validity.} The labels are single-model,
single-pass guesses with no verified ground truth; our evaluation measures
selectivity and distributional conservatism, not precision. Symbol recovery depends on PDB
availability and is incomplete for fully stripped third-party modules.
Reach depth is computed over a statically recovered call graph and misses
indirect and virtual edges, so both reachability tiers and the funnel are
approximations. The priority weights are hand-tuned, not learned.

\paragraph{Future work.} The natural next step is a \emph{calibration loop}: log
human or agent verdicts on sampled candidates, compute true-positive rate by
bug-class, reachability, and module, and use it to re-weight both the classifier
prompt and the priority function, turning a static guess table into a learning
prioritizer, and replacing the hand-tuned weights with a model trained on
verdicts.

\section{Conclusion}
\label{sec:conclusion}

Target selection, not analysis, is a binding constraint when hunting
vulnerabilities across millions of functions, especially as analysis is
delegated to agents that must be pointed somewhere. We showed that a low-cost
\sys\ pipeline produces a selective, queryable prioritization layer over a
production Windows corpus: it reserves its severe labels for a small minority
of functions, demotes the boilerplate floor, grounds its rationales in
verifiable features, and, stacked with deterministic filters, narrows a
millions-strong search space to a tractable shortlist at low per-function
cost. It is not a bug detector and claims no discovered vulnerabilities; it
is the substrate a detector should run on. We describe the methodology and
report aggregate results, and, for legal and dual-use reasons, withhold the
derived dataset. Whether the shortlist improves real bug-finding yield over
uninformed search is the central open question we leave to future work.

\paragraph{Reproducibility.} The pipeline (PDB auto-fetch, call-graph and
feature extraction, structured LLM enrichment, priority-weighted sampler) is
specified in Section~\ref{sec:method}; all inputs are production Windows
binaries publicly retrievable from Microsoft-hosted symbol and file endpoints,
and the figures in this paper are generated from the aggregate counts reported
herein. We do not redistribute any Microsoft binaries, PDBs, function-level
annotations, or sampled queues; only aggregate counts appear in the paper.

\bibliographystyle{plain}
\bibliography{bibtex/references}

@misc{siftrank,
  title  = {Sift or Get Off the {PoC}: Applying Information Retrieval to
            Vulnerability Research with {SiftRank}},
  author = {Gross, Caleb},
  year   = {2025},
  eprint = {2512.06155},
  archivePrefix = {arXiv},
  primaryClass  = {cs.CR},
  note   = {\url{https://arxiv.org/abs/2512.06155}}
}

@misc{onthemoney,
  title  = {{O(N)} the Money: Scaling Vulnerability Research with {LLMs}},
  author = {Gross, Caleb},
  year   = {2025},
  note   = {\url{https://noperator.dev/posts/on-the-money/}}
}

@misc{promptingpriorities,
  title  = {Prompting the Priorities: A First Look at Evaluating {LLMs} for
            Vulnerability Triage and Prioritization},
  author = {Al Haddad, Osama and Ikram, Muhammad and Ahmed, Ejaz and Lee, Young},
  year   = {2025},
  eprint = {2510.18508},
  archivePrefix = {arXiv},
  primaryClass  = {cs.CR},
  note   = {\url{https://arxiv.org/abs/2510.18508}}
}

@inproceedings{needlealpc,
  title     = {Needle in a Haystack: Automated and Scalable Vulnerability
               Hunting in the {Windows} {ALPC} Sea},
  author    = {Liu, Haoyi and Dong, Feng and Tian, Yunpeng and Zhang, Mu and
               Li, Xuefeng and Gu, Fangming and Peng, Zhiniang and Wang, Haoyu},
  booktitle = {Proceedings of the 2025 ACM SIGSAC Conference on Computer and
               Communications Security (CCS)},
  year      = {2025},
  doi       = {10.1145/3719027.3765180},
  note      = {\url{https://doi.org/10.1145/3719027.3765180}}
}

@misc{secvuleval,
  title  = {{SecVulEval}: Benchmarking {LLMs} for Real-World {C/C++}
            Vulnerability Detection},
  author = {Ahmed, Md Basim Uddin and Shiri Harzevili, Nima and Shin, Jiho and
            Pham, Hung Viet and Wang, Song},
  year   = {2025},
  eprint = {2505.19828},
  archivePrefix = {arXiv},
  primaryClass  = {cs.SE},
  note   = {\url{https://arxiv.org/abs/2505.19828}}
}

@misc{llmvulnsurvey,
  title  = {Large Language Model for Vulnerability Detection and Repair:
            Literature Review and the Road Ahead},
  author = {Zhou, Xin and Cao, Sicong and Sun, Xiaobing and Lo, David},
  year   = {2024},
  eprint = {2404.02525},
  archivePrefix = {arXiv},
  primaryClass  = {cs.SE},
  note   = {\url{https://arxiv.org/abs/2404.02525}}
}

@misc{bigsleep,
  title  = {From Naptime to Big Sleep: Using Large Language Models To Catch
            Vulnerabilities In Real-World Code},
  author = {{Google Project Zero (Big Sleep Team)}},
  year   = {2024},
  note   = {\url{https://projectzero.google/2024/10/from-naptime-to-big-sleep.html}}
}

@misc{fang2024oneday,
  title  = {{LLM} Agents can Autonomously Exploit One-day Vulnerabilities},
  author = {Fang, Richard and Bindu, Rohan and Gupta, Akul and Kang, Daniel},
  year   = {2024},
  eprint = {2404.08144},
  archivePrefix = {arXiv},
  primaryClass  = {cs.CR},
  note   = {\url{https://arxiv.org/abs/2404.08144}}
}

@article{heckman2011,
  title   = {A systematic literature review of actionable alert identification
             techniques for automated static code analysis},
  author  = {Heckman, Sarah and Williams, Laurie},
  journal = {Information and Software Technology},
  volume  = {53},
  number  = {4},
  pages   = {363--387},
  year    = {2011},
  doi     = {10.1016/j.infsof.2010.12.007},
  note    = {\url{https://doi.org/10.1016/j.infsof.2010.12.007}}
}

@misc{awisurvey,
  title  = {Machine Learning for Actionable Warning Identification:
            A Comprehensive Survey},
  author = {Ge, Xiuting and Fang, Chunrong and Li, Xuanye and Sun, Weisong and
            Wu, Daoyuan and Zhai, Juan and Lin, Shangwei and Zhao, Zhihong and
            Liu, Yang and Chen, Zhenyu},
  year   = {2024},
  eprint = {2312.00324},
  archivePrefix = {arXiv},
  primaryClass  = {cs.SE},
  note   = {\url{https://arxiv.org/abs/2312.00324}}
}

@inproceedings{aflgo,
  title     = {Directed Greybox Fuzzing},
  author    = {B{\"o}hme, Marcel and Pham, Van-Thuan and
               Nguyen, Manh-Dung and Roychoudhury, Abhik},
  booktitle = {Proceedings of the 2017 ACM SIGSAC Conference on Computer and
               Communications Security (CCS)},
  pages     = {2329--2344},
  year      = {2017},
  doi       = {10.1145/3133956.3134020},
  note      = {\url{https://doi.org/10.1145/3133956.3134020}}
}

@misc{winbindex,
  title  = {{Winbindex}: An Index of {Windows} Binaries},
  author = {{m417z}},
  year   = {2021},
  note   = {\url{https://winbindex.m417z.com/}}
}

@article{efraimidis,
  title   = {Weighted random sampling with a reservoir},
  author  = {Efraimidis, Pavlos S. and Spirakis, Paul G.},
  journal = {Information Processing Letters},
  volume  = {97},
  number  = {5},
  pages   = {181--185},
  year    = {2006},
  doi     = {10.1016/j.ipl.2005.11.003},
  note    = {\url{https://doi.org/10.1016/j.ipl.2005.11.003}}
}

\end{document}